\documentclass[%
reprint,
amsmath,amssymb,
aps,
]{revtex4-2}

\usepackage{graphicx}
\usepackage{dcolumn}
\usepackage{bm}
\usepackage{amsmath,amsfonts,amssymb}
\usepackage{graphicx,epsfig}
\usepackage{color}
\usepackage{captcont}
\usepackage{slashed}
\usepackage{epstopdf}
\usepackage{mathtools}
\usepackage{natbib}
\usepackage{ulem}
\usepackage{subfigure}
\usepackage{graphicx}
\usepackage{CJKutf8}

\begin{document}
	
	\preprint{APS/123-QED}
	
	\title{Optical Appearance of Numerical Black Hole Solutions in Higher Derivative Gravity}
	\author{Yu-Hao Cui$^1$}
	\email{yuhaocui@cug.edu.cn}
	\affiliation{$^1$Hubei Subsurface Multi-scale Imaging Key Laboratory, School of Geophysics and Geomatics, China University of Geosciences, Wuhan 430074, People's Republic of China\\
		$^2$College of Physics and Electronic Engineering, Chongqing Normal University, Chongqing 401331, People's Republic of China\\
		$^3$School of Physics and Astronomy, China West Normal University, Nanchong 637000, People's Republic of China
	}
	\author{Sen Guo$^2$}
	\email{Corresponding author: Sen Guo(sguophys@126.com)}
	\affiliation{$^1$Hubei Subsurface Multi-scale Imaging Key Laboratory, School of Geophysics and Geomatics, China University of Geosciences, Wuhan 430074, People's Republic of China\\
		$^2$College of Physics and Electronic Engineering, Chongqing Normal University, Chongqing 401331, People's Republic of China\\
		$^3$School of Physics and Astronomy, China West Normal University, Nanchong 637000, People's Republic of China
	}
	\author{Yu-Xiang Huang$^3$}
	\email{yxhuangphys@126.com}
	\affiliation{$^1$Hubei Subsurface Multi-scale Imaging Key Laboratory, School of Geophysics and Geomatics, China University of Geosciences, Wuhan 430074, People's Republic of China\\
		$^2$College of Physics and Electronic Engineering, Chongqing Normal University, Chongqing 401331, People's Republic of China\\
		$^3$School of Physics and Astronomy, China West Normal University, Nanchong 637000, People's Republic of China
	}
     \author{Yu Liang$^1$}
    \email{washy2718@outlook.com}
    \affiliation{$^1$Hubei Subsurface Multi-scale Imaging Key Laboratory, School of Geophysics and Geomatics, China University of Geosciences, Wuhan 430074, People's Republic of China\\
    	$^2$College of Physics and Electronic Engineering, Chongqing Normal University, Chongqing 401331, People's Republic of China\\
    	$^3$School of Physics and Astronomy, China West Normal University, Nanchong 637000, People's Republic of China
    }
    \author{Kai Lin$^1$}
    \email{lk314159@hotmail.com}
    \affiliation{$^1$Hubei Subsurface Multi-scale Imaging Key Laboratory, School of Geophysics and Geomatics, China University of Geosciences, Wuhan 430074, People's Republic of China\\
    	$^2$College of Physics and Electronic Engineering, Chongqing Normal University, Chongqing 401331, People's Republic of China\\
    	$^3$School of Physics and Astronomy, China West Normal University, Nanchong 637000, People's Republic of China
    }
    	
\date{\today}
\begin{abstract}
The optical appearance of the numerically black hole solutions within the higher derivative gravity illuminated by an accretion disk context is discussed. We obtain solutions for non-Schwarzschild black holes with $r_0=1$, $r_0=2$, and $r_0=3$. Further analysis of spacetime trajectories reveals properties similar to Schwarzschild black holes, while the $r_0=2$ black hole exhibits significant differences. The results reveal the presence of a repulsive potential barrier for the black hole, allowing only particles with energies exceeding a certain threshold to approach it, providing a unique gravitational scenario for non-Schwarzschild black holes. Additionally, the optical images are derived through numerical simulations by discussing the trajectories of photons in the black hole spacetime. The distribution of radiation flux and the effects of gravitational redshift and Doppler shift on the observed radiation flux are considered. Interestingly, previous analyses of the optical appearance of black holes were conducted within the framework of analytic solutions, whereas the analysis of numerical black hole solutions first appears in our analysis.
\end{abstract}
\maketitle	
\section{INTRODUCTION}
\label{intro}
\par
Einstein’s general relativity (GR) regarded as the standard theory of gravity, but it encounters conflicts with quantum physics in some details. One of the more important issues in theoretical physics is to construct a theory of gravity that is renormalizable. The GR can be understood as an effective theory at low energies. As energy scales increase, higher-order corrections become increasingly significant \cite{1}. One possible solution involves incorporating higher-order curvature terms into the standard Einstein-Hilbert action. By including all possible quadratic curvature terms, Stelle constructed a theory amenable to renormalization, which is the higher derivative gravity \cite{2,3}. The inclusion of a quadratic curvature term comes with the drawback of introducing a non-physical ghost field \cite{4}. Furthermore, in vacuum quadratic gravity within higher derivative gravity, Einstein gravity is equated with some form of stress energy, rendering the Birkhoff theorem invalid \cite{5}. Consequently, theoretically, there exist spherically symmetric solutions beyond the Schwarzschild solutions. Several static spherically symmetric black hole (BH) solutions have been discovered in higher derivative gravity, with discussions on the stability of non-Schwarzschild BHs \cite{6,7,8,9,10}. Additionally, investigations into charged BH solutions and high-precision analytical approximations of numerical solutions have been undertaken \cite{11,12}.
\par
In addition to theoretical investigations, the astronomical detection of real images of black holes holds tremendous interest. The Event Horizon Telescope (EHT) Collaboration published an image of the supermassive object at the center of the Messier(M) 87* galaxy, marking the first direct visual evidence for the existence of BHs \cite{13}. Subsequently, the EHT collaboration released an image of the BH at the center of the Sagittarius (Sgr) A* \cite{14}. The polarized image of M87* BH revealed the structure of magnetic fields and the plasma properties near the BH \cite{15}. These works provide enlightening insights into understanding the matter distribution and the electromagnetic interactions around BHs. 
\par
The BH shadow, caused by light being captured by the BH, stands as one of the most crucial elements in BH images \cite{16}. The shape and size of these shadows are intimately linked to the geometry of the BH spacetime, offering a method for identifying BHs observed through astronomy \cite{17,18,19}.  Grenzebach $et~al$. derived an analytical formula for the shadow of a Kerr–Newman–NUT–(anti-)de Sitter BH, facilitating the convenient visualization of photon regions and shadows \cite{20,20b}. The EHT collaboration's tests on the radiation transmission code of the BH images revealed that the error in the observed radiation flux was merely one percent, suggesting the feasibility of parameter estimation using BH observations \cite{21}. Numerous detailed studies have delved into gravitational lensing and BH shadows within various gravitational spacetime contexts \cite{22,23,24,25,26,27,28,29,30,31,31b,32b,33b,34b}.
\par
In real astrophysical observation, observable BHs are usually encircled by an accretion disk. Gravitational energy released from material within this disk emits a characteristic spectrum of electromagnetic radiation, serving as a light source against the background of the BH \cite{32}. If the accretion rate surpasses that of other objects in the Milky Way, it can manifest as an active galactic nucleus (AGN) theough Eddington accretion. The redshift of iron lines in the spectra of AGN finds solid explanation through BH accretion disks \cite{33}. Research on accretion disk models traces back to the 1970s, where Shakura $et~al$. proposed a geometrical thin, optically thick accretion disk model, also known as the standard model \cite{34}. Thorne $et~al$. extended this standard model to relativity, establishing the Novikov-Thorne model \cite{35}. Buding upon these studies, Luminet obtained direct and secondary images of the disk of a Schwarzschild BH with a thin accretion disk, calculating the accretion disk's brightness using the radiation flux analytical formula derived in \cite{36}. Utilizing Luminet's approach, images of BHs and naked singularities within various modified gravitis have been investigated \cite{37,38,39,40,41}. 
\par
In addition to optically thick, geometrically thin accretion disks, discussions also encopass other types of accretion disks. Gralla $et~al$. provided a precise description of shadows, photon rings, and lensing rings for optically thin disk \cite{42}. Pedro $et~al$. discussed the lensing and shadow of a BH surrounded by a heavy accretion disk \cite{43}. The cornerstone of these works lies in ray-tracing techniques. However, Luminet's ray-tracing method is only applicable to spherically symmetric spacetime. For axially symmetric BHs, the Newman-Janis algorithm can be utilized to reduce the second-order geodesic equations to a set of first-order geodesic equations by means of a constant of motion \cite{44}. However, in cases where the dynamical system of photons is nonintegrable \cite{45,46,47,48,49}, resulting in equations of motion lacking a carter constant, the motion of the photons is chaotic, and the disk image can only be obtained through a numerical ray-tracing code \cite{48}.

\par
Research on the optical appearance of BHs in the past has been conducted in the context of analytical solutions to BH, which typically apply to simplified BH models. In real astronomical observations, the considered physical backgrounds are complex, involving asymmetry, rotation, or the combination of gravitational fields with other matter, which severely restricts the applicability of analytical solutions. The introduction of numerical BH solutions helps fill this gap, allowing for a more accurate test of the validity of GR. Therefore, the main goal of this article is to discuss the observable features of BH shadows under the backdrop of numerical solutions. The structure of this paper is as follows: in Sec. \ref{sec:2}, we briefly review the non-Schwarzschild BH numerical solutions in higher derivative gravity. Sec. \ref{sec:3} analyzes the time-like geodesics and null geodesics of non-Schwarzschild BH. In Sec. \ref{sec:4}, we explore the optical appearance of this intriguing numerical BH solution within the framework of a thin accretion disk model, while considering the influences of redshift and observed inclination. Finally, we draw our conclusions in Sec. \ref{sec:5}.
	
	
\section{NUMERICAL BLACK HOLE SOLUTION}
\label{sec:2}
\par
The most general Einstein-Hilbert action with additional quadratic curvature terms can be expressed as \cite{6} 
\begin{equation} \label{eq1}
I=d^4x\sqrt{-g}(\gamma R-\alpha C_{\rm \mu \nu \rho \sigma}C^{\rm \mu \nu \rho \sigma}+\beta R^2),
\end{equation}
where $C_{\rm \mu \nu \rho \sigma}$ represents the Weyl tensor, and $\alpha$, $\beta$, and $\gamma$ denote coupling constants. Since the resulting tensor in the field equations arising from the Weyl tensor is traceless, the equations of motion should exclude the contribution of the $R^2$ term . Therefore, it is feasible to assign $\beta=0$. Assuming that $\gamma=1$ to streamline the calculation, the equations of motion can be simplified as \cite{6} 
\begin{equation}\label{eq2}
R_{\mu \nu}-\frac{1}{2}g_{\mu \nu}R-4\alpha B_{\mu \nu}=0,
\end{equation}
where the expression $B_{\rm \mu \nu}=(\nabla^\rho \nabla^\sigma+\frac{1}{2}R^{\rho \sigma}) C_{\rm \mu \rho \nu \sigma}$ defines the Bach tensor. It is notable that the Schwarzschild solution satisfies Eq.\ (\ref{eq2}). A numerical solution for a non-Schwarzschild BH with static spherical symmetry was presented in \cite{6}. The metric for a static spherically symmetric spacetime can be represented as:
\begin{equation}\label{eq3}
ds^2=-h(r)dt^2+\frac{dr^2}{f(r)}+r^2(d\theta^2+\sin^2\theta d \varphi^2).
\end{equation}

\par
Substituting Eq.\ (\ref{eq3}) into Eq.\ (\ref{eq2}), one obtains the field equations:
\begin{equation}\label{eq4}
\begin{split}
rh[rf'h'+2f(rh''+2h')]\\+4h^2(rf'+f-1)-r^2fh'^2=0,
\end{split}
\end{equation}
\begin{equation}\label{eq5}
\begin{split}
f''+\frac{r^2fh'^2+2fhh'+4(f-1)h^2}{2rfh(rh'-2h)}f'\\-\frac{3hf'^2}{4fh-2rfh'} +\frac{r^3fh'+(r^2f-r^2)h}{\alpha r^2f(rh'-2h)}\\+\frac{r^3fh'^3-3r^2fhh'^2-8(f-1)h^3}{2r^2h^2(rh'-2h)}=0.
\end{split}
\end{equation}

\par
Obtaining an analytical solution of Eq. (\ref{eq4}) and Eq. (\ref{eq5}) is challenging, necessitating a numerical approach for solution. We postulate the presence of a BH event horizon at $r=r_0>0$, where $h(r)=f(r)=0$. Near this horizon, one can derive Taylor expansions for $f(r)$ and $h(r)$:
\begin{equation}\label{eq6}
h(r)=c[(r-r_0)+h_2(r-r_0)^2+h_3(r-r_0)^3+\cdots],
\end{equation}
\begin{equation}\label{eq7}
f(r)=f_1(r-r_0)+f_2(r-r_0)^2+f_3(r-r_0)^3+\cdots,
\end{equation}
in which $c$ can be absorbed into a rescaling of the time coordinate, and $f_i$ and $h_i$ denote the constant coefficients characterizing $h(r)$ and $f(r)$ in the vicinity of the event horizon. We designate $h_1=f_1$ to accommodate the arbitrary rescaling of the time coordinate. Setting $\alpha=\frac{1}{2}$ without loss of generality allows for the calculation of all coefficients $h_i$ and $f_i$ with $i>2$ from two nontrivial free parameters, namely $r_0$ and $f_1$. The parameter values are presented in Table \ref{tt11}. Utilizing numerical routines in MATHEMATICA, we integrate the equations. It's crucial to emphasize that $r_0$ needs to be determined first, and subsequently, a specific $f_1$ is obtained through a shooting method to ensure the integration remains free of singularities over a large radius.

\begin{table}
\caption{The numerical value of $f_i$ and $h_i$ ($i\leq 3$). }
\label{tt11}
\begin{ruledtabular}
\begin{tabular}{cccccc}
\\[-7pt]
$r_0$ & $f_1$ & $f_2$ & $f_3$ & $h_2$ & $h_3$\\
\\[-7pt]
\hline
\\[-6pt]
1 & 1.36330 & -1.52674 & -1.60471 & -1.79323 & 2.11397\\[1pt]
2 & 3.04233 & -2.1655 & 1.25404 & -3.00115 & 2.18269\\[1pt]
3 & 4.65416 & -2.29528 & 0.89807 & -3.22376 & -1.60282
\end{tabular}
\end{ruledtabular}
\end{table}

\begin{figure*}
\centering
\begin{minipage}{0.3\textwidth}
\centering
\includegraphics[width=1\textwidth]{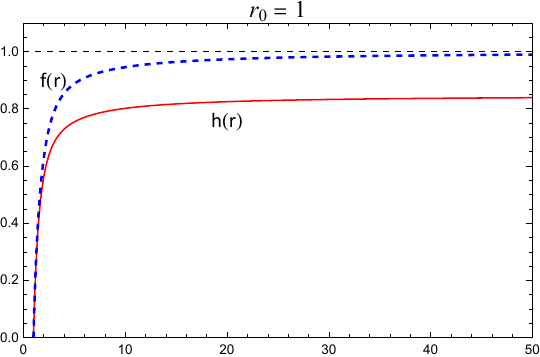}
\end{minipage}
\begin{minipage}{0.3\textwidth}
\centering
\includegraphics[width=1\textwidth]{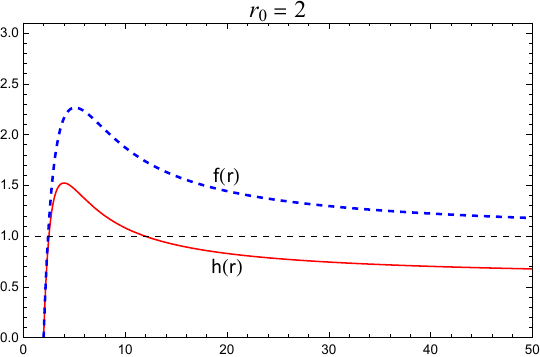}
\end{minipage}
\begin{minipage}{0.3\textwidth}
\centering
\includegraphics[width=1\textwidth]{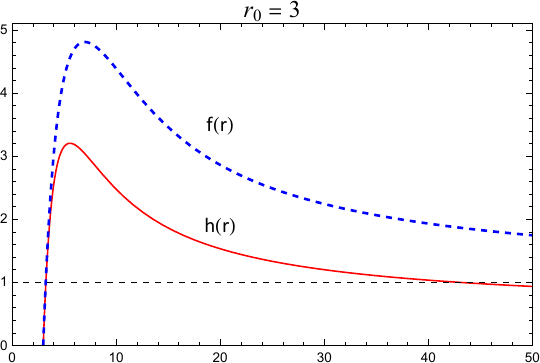}
\end{minipage}
\caption{Numerical solutions for $f(r),h(r)$ in the case of $r_0=1$, $r_0=2$ and $r_0=3$. The dashed black line represents the unity. In each plot, We set $c=\frac{3}{4}$ to separate $h(r)$ from $f(r)$.}
\label{t1}
\end{figure*}	

\par
Figure \ref{t1} illustrates the behavior of $h(r)$ and $f(r)$ for various values of $r_0$. In the analysis by \cite{6}, a critical event horizon radius $r_c\approx 1.143$ is identified. For $r_0 < r_c$, both $h(r)$ and $f(r)$ are increasing functions, while for $r_0 > r_c$, $h(r)$ and $f(r)$ display extremal points. In both two cases, the metric function $f(r)$ will approach $1$ when $r\rightarrow{+\infty}$, so the non-Schwarzschild solutions are asymptotically flat. Furthermore, the mass of the non-Schwartzschild black hole decreases as $r_0$ increases, and becoming negative when $r_0>r_c$ \cite{6}. We will examine the class of $r_0<r_c$ and $r_0>r_c$ by taking $r_0$ as 1 and 2 respectively.

\section{GEODESICS OF HIGHER DERIVATIVE GRAVITY BLACK HOLE}
\label{sec:3}
\subsection{Time-like Geodesic}
By leveraging the spherical symmetry of spacetime, we confine the particle motion to the equatorial plane ($\theta=\frac{\pi}{2}$) without sacrificing generality. As the metric lacks explicit dependence on the coordinates $t$ and $\varphi$, two conserved quantities governing the motion emerge:
\begin{equation}\label{eq8}
p_t=g_{tt}\frac{dt}{d\tau}=E,
\end{equation}
\begin{equation}\label{eq9}  
p_\varphi=g_{\varphi \varphi}\frac{d\varphi}{d\tau}=L,
\end{equation}
where $\tau$ represents the proper time of the particle, while $E$ and $L$ denote the energy and angular momentum, respectively. The time-like geodesic is governed by $ds^2=-d\tau^2$, By Substituting Eq.\ (\ref{eq8}) and Eq.\ (\ref{eq9}) into the geodesic equation, one can derive the equations governing radial motion and orbital dynamics for particles:
\begin{equation}\label{eq10}
\centering
\left(\frac{dr}{d\tau}\right)^2=f(r)\left(\frac{E^2}{h(r)}-\frac{L^2}{r^2}-1\right),
\end{equation}
\begin{equation}\label{eq11}
\centering
\left(\frac{dr}{d\varphi}\right)^2=r^4f(r)\left(\frac{E^2}{L^2h(r)}-\frac{1}{r^2}-\frac{1}{L^2}\right).
\end{equation}

\par
Taking the derivative of both sides of Eq. (\ref{eq10}) with respect to $\tau$, one obtains the expression for radial acceleration:
\begin{equation}\label{eq12}
\centering
\begin{split}
2\frac{d^2r}{d\tau^2}=f'(r)\left(\frac{E^2}{h(r)}-\frac{L^2}{r^2}-1\right)\\+f(r)\left[\frac{2L^2}{r^3}-\frac{E^2h'(r)}{h(r)^2}\right].
\end{split}
\end{equation}

\par
Considering the orbit of the particle on the accretion disk as a circular orbit, we initially examine the scenario of such orbit. For particles in a circular orbit, both radial velocity and acceleration are zero ($\dot{r}=\ddot{r}=0$). Hence, one can derive expressions for $E$ and $L$ in terms of $r$:
\begin{equation}\label{eq13}
E(r)^2=\frac{2h(r)}{2h(r)-rh'(r)},\ \ L(r)^2=\frac{r^3h'(r)}{2h(r)-rh'(r)}.
\end{equation}

\par
As energy and angular momentum must be real numbers, the conditions for a circular orbit are $2h(r)-rh'(r)>0$ and $h'(r)>0$. Utilizing the numerical values of BH calculated in Sec.\ \ref{sec:2}, one can determine the range of circular orbits for non-Schwarzschild BH, i.e,
\begin{equation}\label{eq14}
r_{r_0=1}>1.437,\ \ \ 2.666<r_{r_0=2}<3.996.
\end{equation}
\begin{figure*}
\centering
\begin{minipage}{0.46\textwidth}
\centering
\includegraphics[width=1\textwidth]{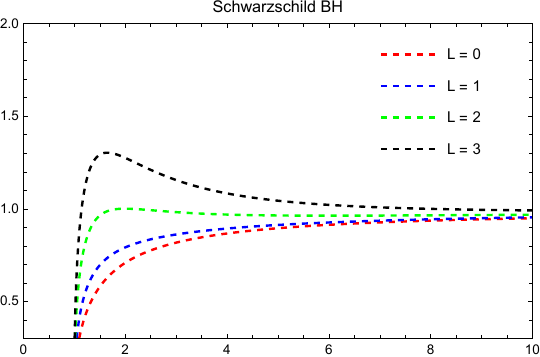}
\end{minipage}
\begin{minipage}{0.46\textwidth}
\centering
\includegraphics[width=1\textwidth]{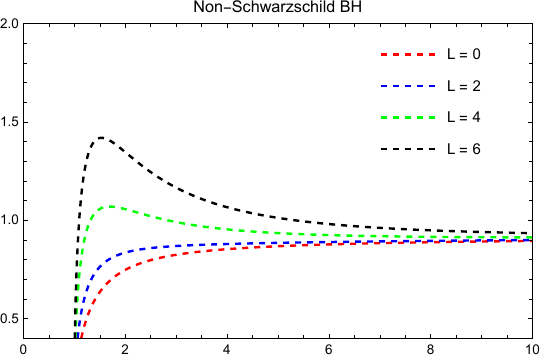}
\end{minipage}
\begin{minipage}{0.46\textwidth}
\centering
\includegraphics[width=1\textwidth]{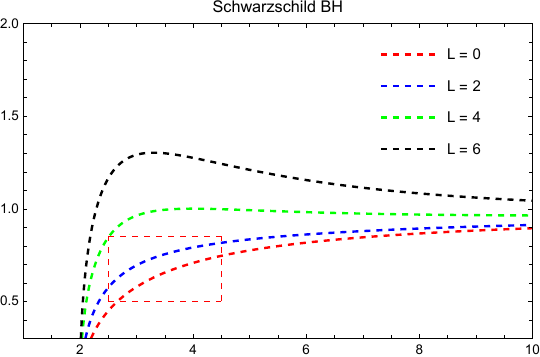}
\end{minipage}
\begin{minipage}{0.46\textwidth}
\centering
\includegraphics[width=1\textwidth]{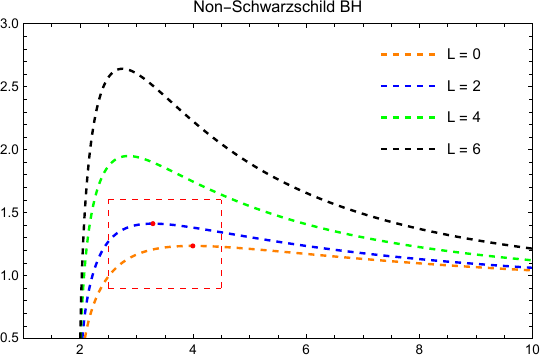}
\end{minipage}
\caption{Effective potential as a function of $r$ for two types of BHs. Top panle: $r_0=1$. Bottom panle: $r_0=2$}
\label{t2}
\end{figure*}
	
\par
Moreover, the stability of a circular orbit necessitates $V''(r)\leq0$. Our calculations for non-Schwarzschild BH reveal that the radius of the inner stable circular orbit is approximately $r_{isco}\approx3.503$. However, in cases where $r_0>r_c$, our calculations demonstrate that $V''(r)>0$ within the range of conditions for circular orbit. This indicates the absence of unstable circular orbits for BH with $r_0>r_c$. For general particle orbits, to simplify the description of radial motion, we introduce the particle's effective potential $V_{eff}$, defined as the energy with zero radial velocity. This potential can be derived from Eq. (\ref{eq10})
\begin{equation}\label{eq15}
V_{eff}=h(r)(\frac{L^2}{r^2}+1)
\end{equation}
	
\par
Figure \ref{t2} illustrates the radial distribution of the effective potential for two types of BHs with $r_0=1$ and $r_0=2$. In the case of $r_0=1$, there is minimal distinction between the Schwarzschild BH and the non-Schwarzschild BH counterpart. For the Schwarzschild BH, extreme points in the effective potential emerge only when $L>\sqrt{3}$. This implies that particles with low angular momentum moving towards the BH are inevitably captured, while those with high angular momentum must overcome a potential barrier to approach the BH. However, in the case of $r_0=2$, the non-Schwarzschild BH exhibits a distinct behavior. Even when angular momentum is negligible, the potential function still features a unique extreme value $V_{max}$. Consequently, particle with any angular momentum outside the corresponding radius of this extreme point must traverse a potential barrier to approach the BH. 

\par
\begin{figure}
\centering
\begin{minipage}{0.23\textwidth}
\centering
\includegraphics[width=1\textwidth]{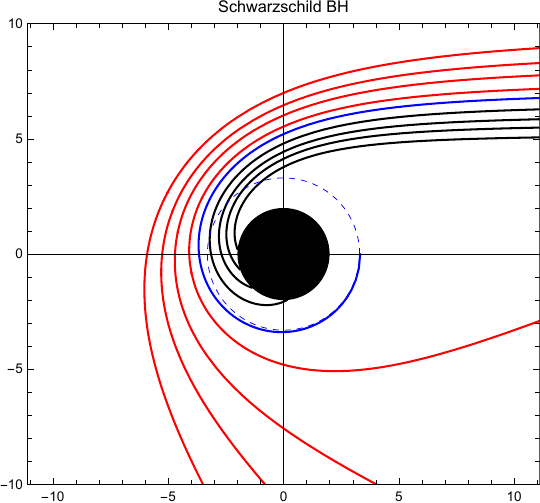}
\end{minipage}
\begin{minipage}{0.23\textwidth}
\centering
\includegraphics[width=1\textwidth]{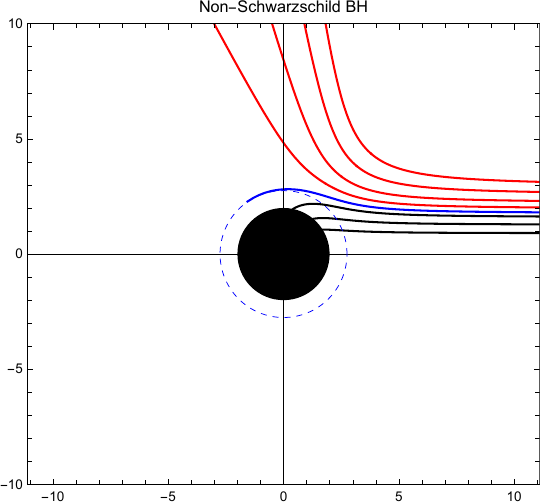}
\end{minipage}
\caption{The particle trajectories of two types of BHs  with $r_0=2$. The BHs are shown as the black disks, and the dashed blue lines represent the circular orbit.}
\label{t3}
\end{figure}

To illustrate the contrast between the orbits of two types of BHs with $r_0=2$, Fig.\ \ref{t3} depicts the trajectories of unbound particles with varying energies around the BHs, assuming $L=6$. The plot reveals the repulsive effect exerted by the non-Schwarzschild BH on distant particles. In particular, if the radial velocity of the particle is 0, Eq. (\ref{eq12}) can be reduced to
\begin{equation}\label{eq16}
\frac{d^2r}{dt^2}=-\frac{f(r)[r^3h'(r)+L^2(-2h(r)+rh'(r))]}{2r^3h(r)}.
\end{equation}	
	
\par
The sign on the right side of Eq.\ (\ref{eq16}) denotes the direction of the BH force. It's evident that $\ddot{r}>0$ holds true whenever $h'(r)<0$, indicating the effective repulsive nature of the BH on particles. For a particle in a circular orbit, its radial velocity is zero, hence its acceleration can be described by Eq.\ (\ref{eq16}). This elucidates why there exists an upper limit for the orbit radius of the BHs with $r_0=2$ and $r_0=3$ in Eq.\ (\ref{eq14}). Beyond this limit, the force from the BH is repulsive, thereby preventing the formation of circular orbits.
	
\subsection{Null Geodesic}
In our earlier examination, we determined that a non-Schwarzschild BH with $r_0> r_c$ lacks a stable circular orbit, thereby precluding the formation of a stable accretion disk. Consequently, our discussion regarding the accretion disk's image solely focuses on the scenario where $r_0=1$. Analogous to the scrutiny of time-like geodesics, we delve into the trajectory of photons within the equatorial plane:
\begin{equation}\label{eq17}
p_t=g_{tt}\frac{dt}{d\lambda}=E,
\end{equation}
\begin{equation}\label{eq18}  
p_\varphi=g_{\varphi \varphi}\frac{d\varphi}{d\lambda}=L,
\end{equation}
where $\lambda$ represents the affine parameter, $E$ denotes the energy, and $L$ signifies the angular momentum of the photon. The null geodesic adheres to the condition $ds^2=0$. Following the approach employed in defining the effective potential for particles in the preceding section, we define the effective potential for photons as:
\begin{equation}\label{eq19}
V_{eff}=h(r)\frac{L^2}{r^2}.
\end{equation}

\par
For photons, the angular momentum $L$ solely influences the magnitude of the effective potential, without altering its behavior. The radius of the photon sphere is determined at points where $V_{eff}=0$ and $V_{eff}'=0$. Hence, One can derive the radius of the photon sphere sphere and the corresponding impact parameters by:
\begin{equation}\label{eq20}
Schwarzschild\ BH\ \ \ \ \ \ \ \ \ \ r_{s}=1.5\ \  b_{c}=\frac{3\sqrt{3}}{2},
\end{equation}
\begin{equation}\label{eq21}
Non-Schwarzschild\ BH\ \ \ r_{s}\approx1.437 \ \ b_{c}\approx2.356,
\end{equation}
where $b=\frac{L}{E}$ represents the impact parameter for photons at infinity. Subsequently, incorporating Eq. (\ref{eq17}) and Eq. (\ref{eq18}) into $ds^2=0$, and eliminating $\tau$ from the equation, yields the orbital equation:
\begin{equation}\label{eq22}
\left(\frac{dr}{d\varphi}\right)^2=r^4f(r)\left(\frac{1}{b^2h(r)}-\frac{1}{r^2}\right).
\end{equation}
	
\par
\begin{figure}
\centering
\begin{minipage}{0.23\textwidth}
\centering
\includegraphics[width=1\textwidth]{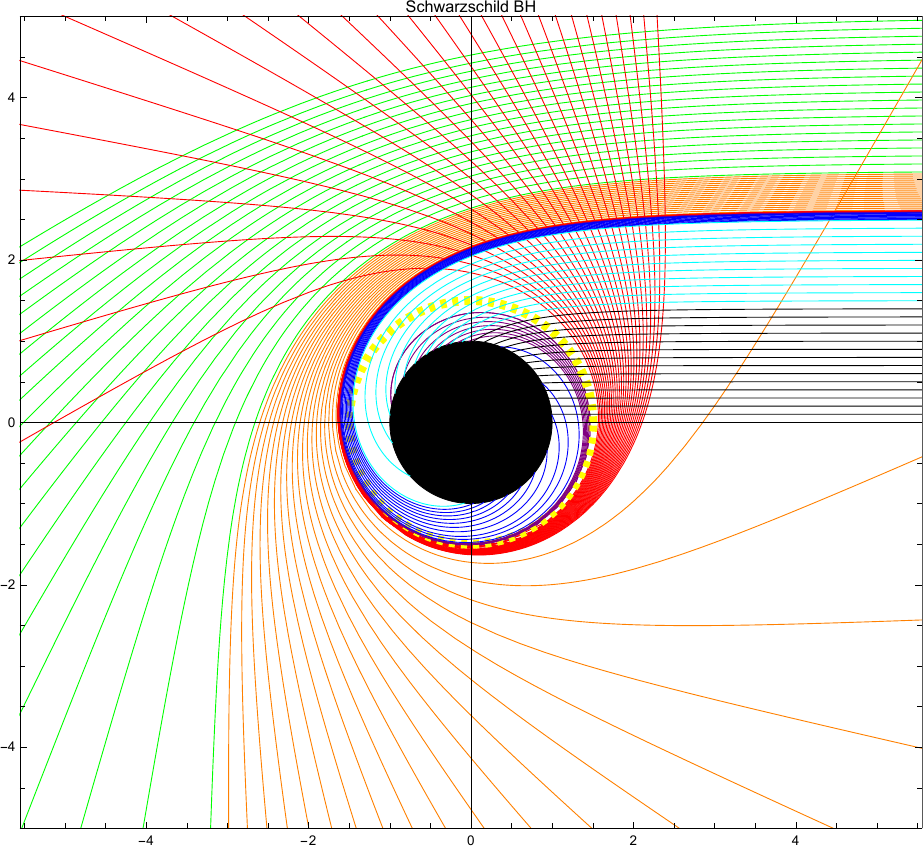}
\end{minipage}
\begin{minipage}{0.23\textwidth}
\centering
\includegraphics[width=1\textwidth]{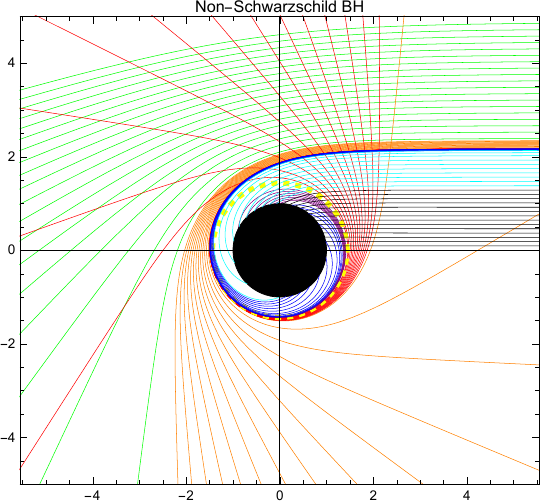}
\end{minipage}
\caption{The light trajectories of two types of BHs with $r_0=1$, the BHs are shown as the black disks, and the yellow dashed circle represent the circular orbit.}
\label{t4}
\end{figure}

To obtain the deflection angle of light, integrate $r$ in the orbital equation (\ref{eq22}):
\begin{equation}\label{eq23}
\gamma_1=\int_{r_{sourse}}^{+\infty}\frac{dr}{\sqrt{r^4f(r)(\frac{1}{b^2h(r)}-\frac{1}{r^2})}}.
\end{equation}
In this equation, $\gamma_1$ denotes the change in the $\varphi$-coordinate of a photon from the radius $r_{\text{sourse}}$ to infinity. It's worth noting that for the photon passing the perihelion, the expression for the deflection angle is:
\begin{equation}\label{eq24}
\begin{split}
\gamma_2=2\int_{r_p}^{+\infty}\frac{dr}{\sqrt{r^4f(r)(\frac{1}{b^2h(r)}-\frac{1}{r^2})}}\\-\int_{r_{sourse}}^{+\infty}\frac{dr}{\sqrt{r^4f(r)(\frac{1}{b^2h(r)}-\frac{1}{r^2})}}
\end{split}
\end{equation}
where $r_p$ represents the radius at perihelion. Fig. \ref{t4} illustrates the photon orbit of two types of BHs. For $b>b_c$, the green line corresponds to direct image, the orange line corresponds to lensed ring and the red line corresponds to photon ring. For $b>b_c$, The cyan line, blue line, purple line represent the direct image, lens ring, and photon ring, respectively. \cite{50}
It's observable that light exhibiting the same deflection around non-Schwarzschild BHs possesses slightly smaller impact parameters.

\begin{figure}
\centering
\begin{minipage}{0.46\textwidth}
\centering
\includegraphics[width=1\textwidth]{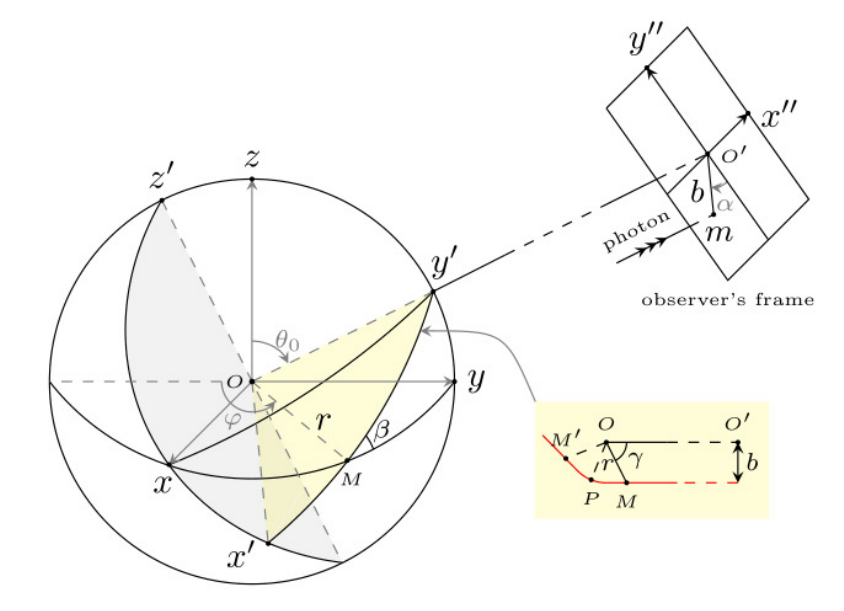}
\end{minipage}
\caption{The schematic diagram is indicated in Ref. \cite{36}.}
\label{t55}
\end{figure}

\begin{figure*}
\centering
\begin{minipage}{0.3\textwidth}
\centering
\includegraphics[width=1\textwidth]{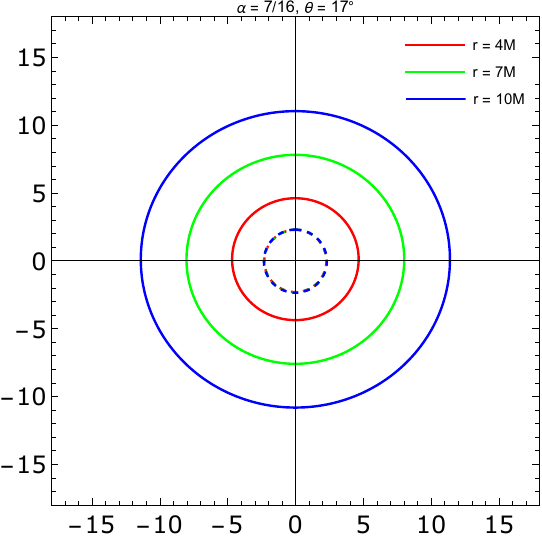}
\end{minipage}
\begin{minipage}{0.3\textwidth}
\centering
\includegraphics[width=1\textwidth]{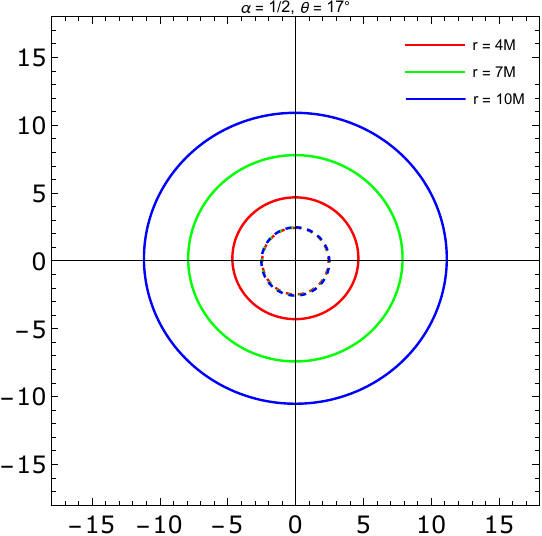}
\end{minipage}
\begin{minipage}{0.3\textwidth}
\centering
\includegraphics[width=1\textwidth]{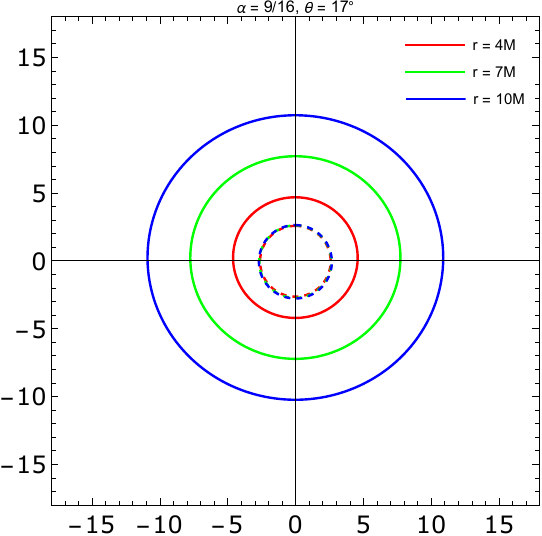}
\end{minipage}
\begin{minipage}{0.3\textwidth}
\centering
\includegraphics[width=1\textwidth]{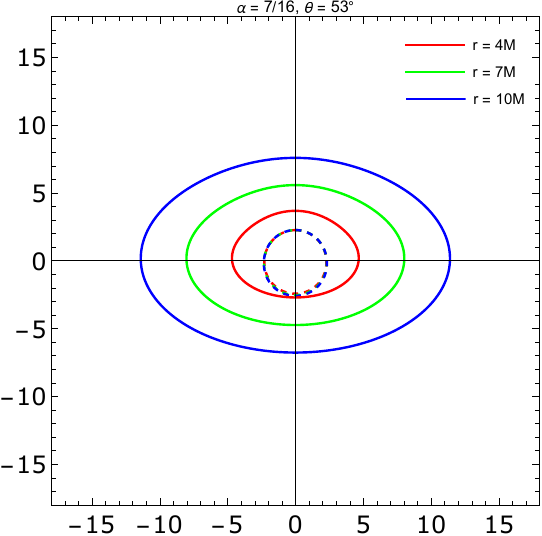}
\end{minipage}
\begin{minipage}{0.3\textwidth}
\centering
\includegraphics[width=1\textwidth]{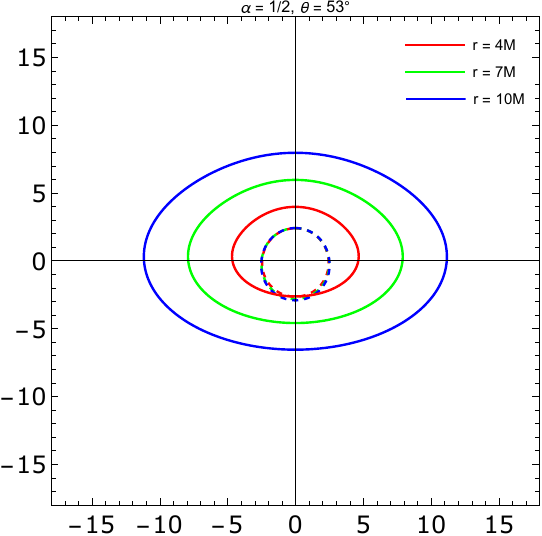}
\end{minipage}
\begin{minipage}{0.3\textwidth}
\centering
\includegraphics[width=1\textwidth]{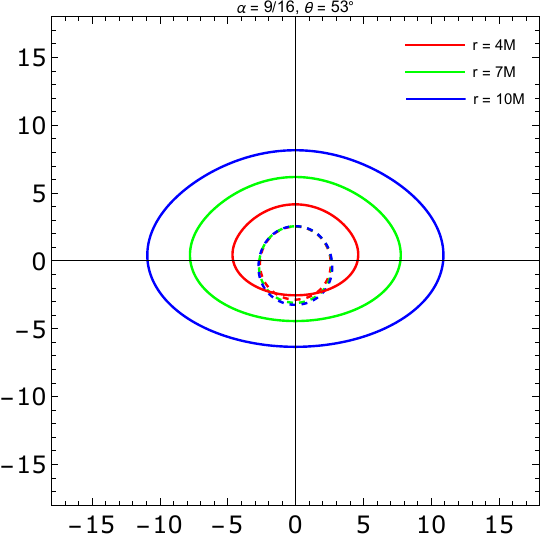}
\end{minipage}
\begin{minipage}{0.3\textwidth}
\centering
\includegraphics[width=1\textwidth]{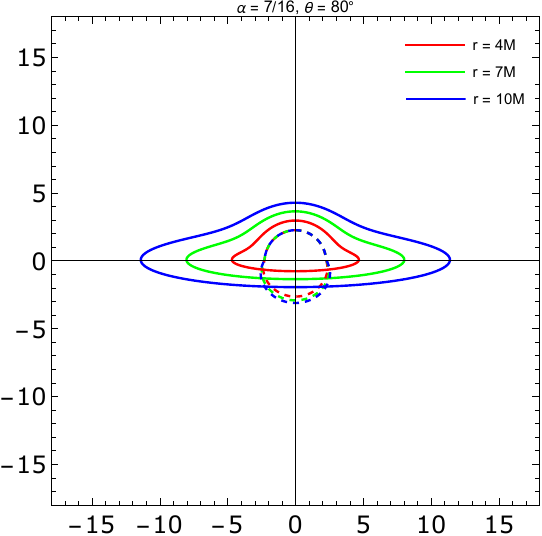}
\end{minipage}
\begin{minipage}{0.3\textwidth}
\centering
\includegraphics[width=1\textwidth]{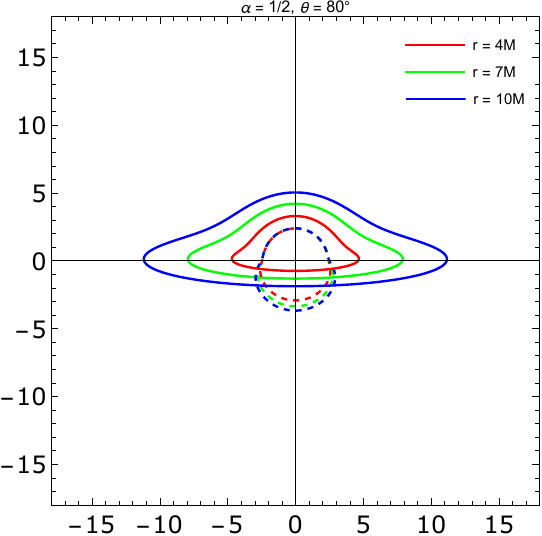}
\end{minipage}
\begin{minipage}{0.3\textwidth}
\centering
\includegraphics[width=1\textwidth]{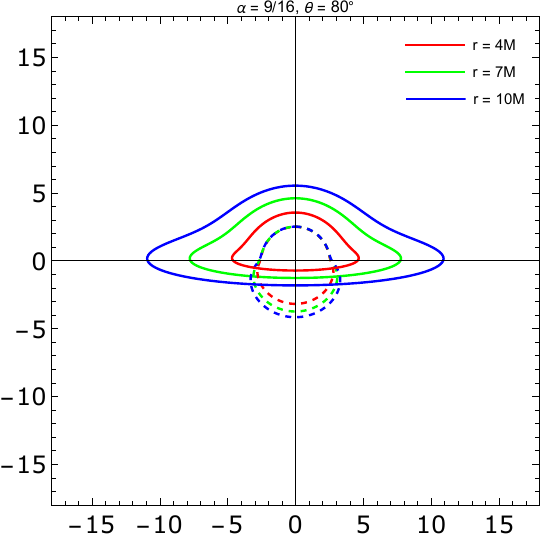}
\end{minipage}
\caption{The direct (solid line) and secondary (dashed line) of non-Schwarzschild BHs with the viewing angles $\theta_0=17^\circ$, $53^\circ$ and $80^\circ$. Left panel: $\alpha=\frac{7}{16}$. Middle panel: $\alpha=\frac{1}{2}$. Right panel: $\alpha=\frac{9}{16}$.}
\label{t66}
\end{figure*}

\section{OBSERVABLE FEATURES OF THE THIN ACCRETION DISK}
\label{sec:4}
\par
In this section, we consider a scenario where the BH is encircled by a geometrically thin, optically thick accretion disk. We utilize the ray-tracing method outlined in \cite{36} to explore the direct and secondary images of the BH. The schematic diagram of the ray-tracing method is depicted in Fig. \ref{t55}, with detailed geometric relationships elaborated in \cite{49}. Here, the BH is situated at point $o$, the accretion disk lies in the $xoy$ plane, the photographic plate at infinity is positioned in the $x''o'y''$ plane with an inclination angle $\theta_0$, and light emanates from point $M$ ($r,\varphi$) on the accretion disk, arriving at point $m$ on the observation plane in the progressive direction $oo'$. Our primary focus lies on the direct image ($b^{(d)},\alpha$) generated by photons facing the photographic plate and the secondary image ($b^{(s)},\alpha+\pi$) generated by photons facing away from the photographic plate.
\begin{figure*}
\centering
\begin{minipage}{0.3\textwidth}
\centering
\includegraphics[width=1\textwidth]{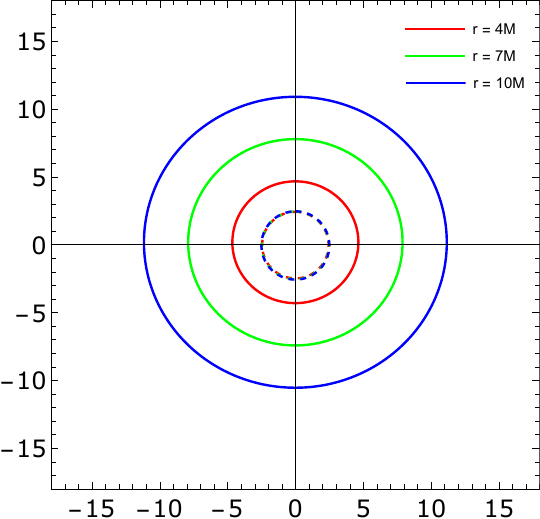}
\end{minipage}
\begin{minipage}{0.3\textwidth}
\centering
\includegraphics[width=1\textwidth]{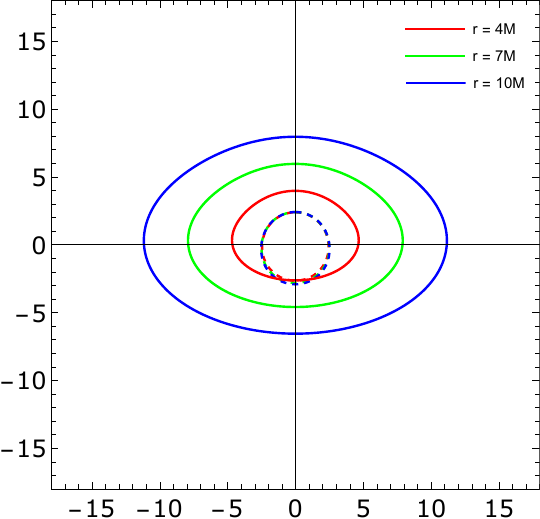}
\end{minipage}
\begin{minipage}{0.3\textwidth}
\centering
\includegraphics[width=1\textwidth]{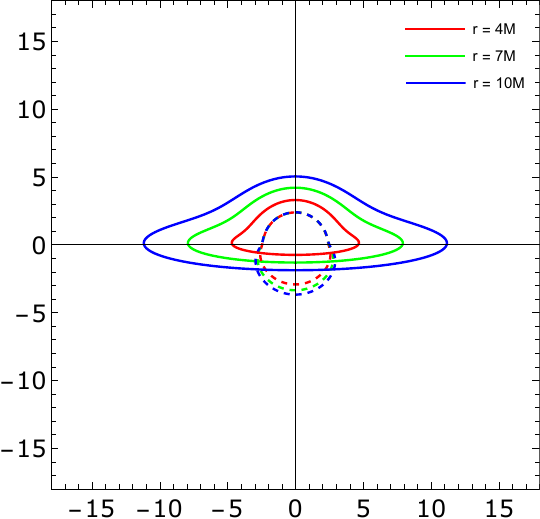}
\end{minipage}
\begin{minipage}{0.3\textwidth}
\centering
\includegraphics[width=1\textwidth]{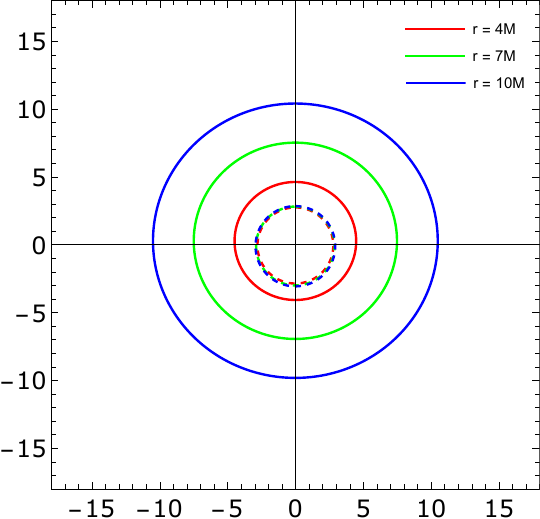}
\end{minipage}
\begin{minipage}{0.3\textwidth}
\centering
\includegraphics[width=1\textwidth]{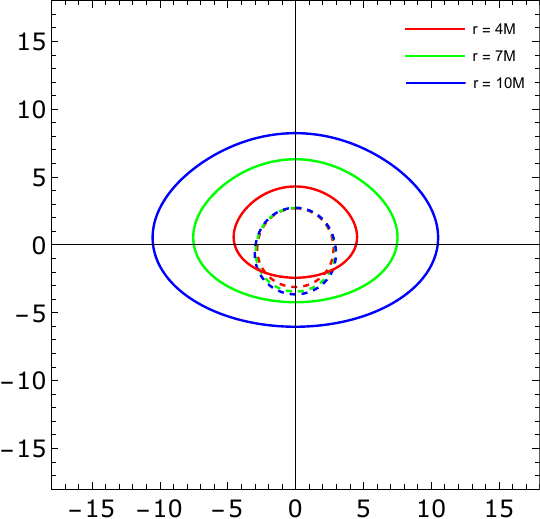}
\end{minipage}
\begin{minipage}{0.3\textwidth}
\centering
\includegraphics[width=1\textwidth]{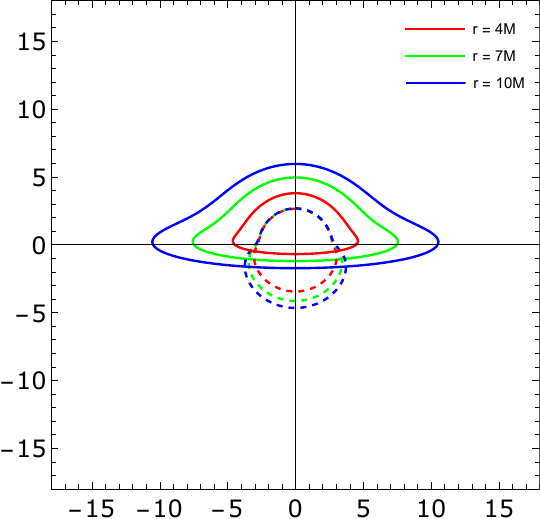}
\end{minipage}
\caption{The direct (solid line) and secondary (dashed line) of two types of BHs with the viewing angles $\theta_0=17^\circ$, $53^\circ$ and $80^\circ$. Top panel: Non-Schwarzschild BH. Bottom panel: Schwarzschild BH.}
\label{t7}
\end{figure*}

\par
\subsection{Ray-tracing}
As depicted in Fig. \ref{t55}, the photon traverses the $oo'M$ plane, and the deflection angle $\gamma$ precisely corresponds to $\angle Moo'$. By applying the sine theorem of spherical triangles to $\triangle Myy'$ and $\triangle Mxx'$, one can derive:
\begin{equation}\label{eq25}
\cos{\alpha}=\cos{\gamma}\sqrt{\cos^2{\alpha}+\cot^2{\theta_0 }}.
\end{equation}
	
\par
In Section \ref{sec:3}, we utilized numerical integration to illustrate the relationship between the deflection angle and the impact parameter $b$ as outlined in Eq. (\ref{eq23}) and Eq. (\ref{eq24}). In Luminet's analysis, the deflection angle is expressed through elliptic integration. However, our calculations concerning Schwarzschild BHs demonstrate that both numerical and elliptic integration yield nearly identical results within a highly precise range. Hence, we employ the numerical integration method to compute the deflection angle for non-Schwarzschild BHs. For photons corresponding to the direct image and higher-order images, one can derive from \cite{37}:
\begin{equation}\label{eq26}
\gamma=\gamma_1\ \ or\ \ \gamma_2,
\end{equation}
\begin{equation}\label{eq27}
2n\pi-\gamma=\gamma_1\ \ or\ \ \gamma_2,
\end{equation}
in which the parameter $n$ signifies the order of the image. Given that higher-order images closely resemble secondary images, we solely focus on secondary images where $n=1$ \cite{37}. 

\par
In Section \ref{sec:2}, we calculate the non-Schwarzschild solution for $\alpha=\frac{1}{2}$. Despite $\alpha$ and $r_0$ are not independent \cite{6}, we can still obtain a solution with $r_0=1$ in a small neighborhood around $\alpha=\frac{1}{2}$. Figure \ref{t66} shows the direct and secondary images of non-Schwarzschild BH with $\alpha=\frac{7}{16}$, $\alpha=\frac{1}{2}$ and $\alpha=\frac{9}{16}$, We can see that as $\alpha$ increases, the secondary image gradually becomes larger, while the direct image changes slightly. For the sake of convenience in calculations, we still consider the case of $\alpha=\frac{1}{2}$ to investigate the differences between non-Schwarzschild BHs and Schwarzschild BHs.

\par
Figure \ref{t7} illustrates the direct and secondary images of two types of BHs. It's evident that  the observed viewing angle profoundly influences both the direct and secondary images. When the viewing angle is small, the secondary image resembles the direct image, forming a structure akin to the "photon ring". As the viewing angle increases, the direct image assumes a "hat" shape, and the secondary image gradually separate. We observed that while the direct images of the two types of BHs exhibit minimal disparity, the secondary images of non-Schwarzschild BH are notably less separated compared to Schwarzschild BH.
	
\subsection{Observed Flux}
	
\begin{figure*}[htbp]
\centering
\includegraphics[width=18.2cm,height=10cm]{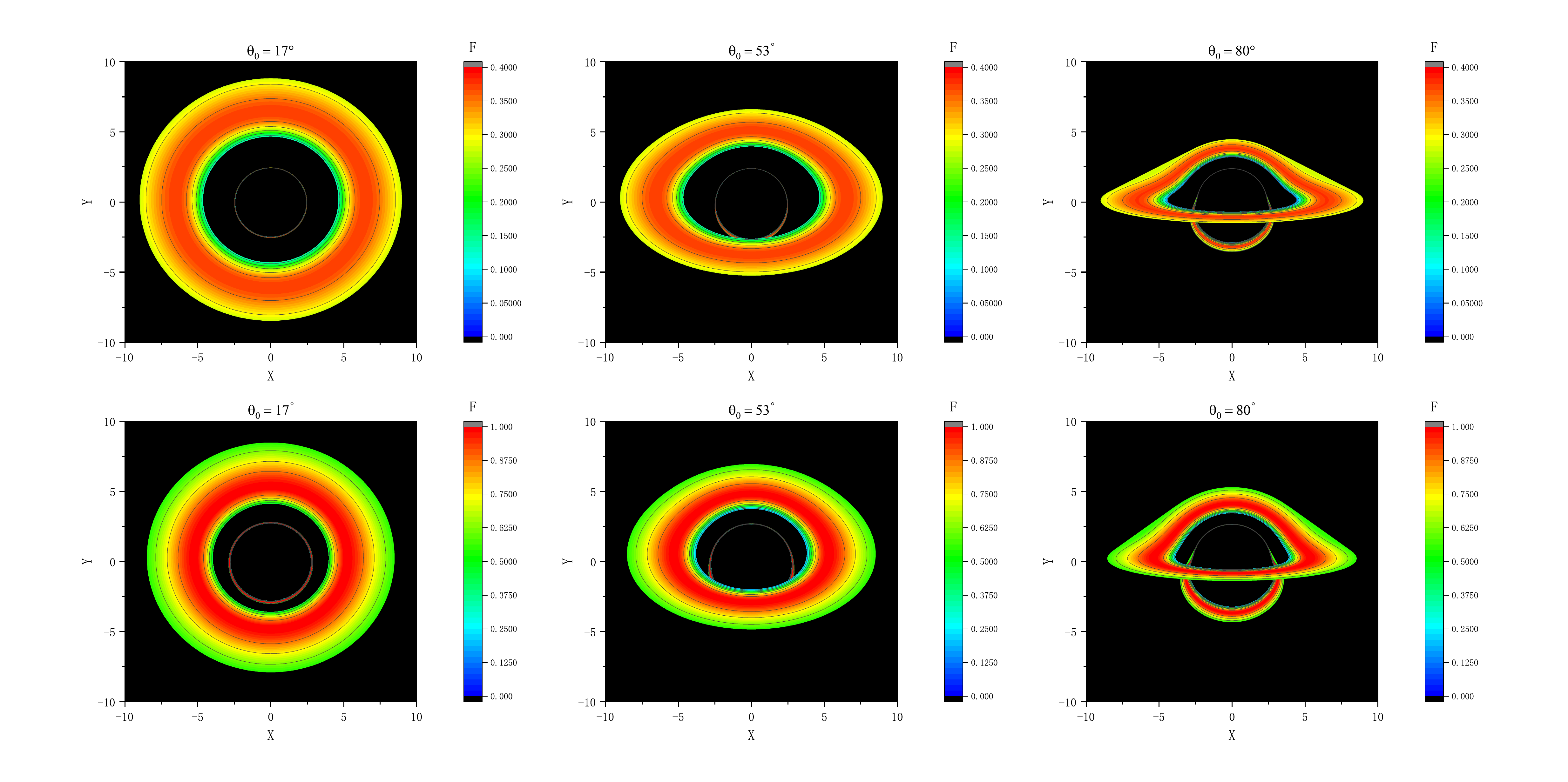}
\caption {The intrinsic flux distribution of two types of BHs with the viewing angles $\theta_0=17^\circ$, $53^\circ$ and $80^\circ$. Top panel: Non-Schwarzschild BH. Bottom panel: Schwarzschild BH.}
\label{t8}
\end{figure*}
	
\par
In this subsection, we will employ the Novikov-Thorne model to analyze the radiation flux of the accretion disk. The expression for the radiation flux is provided in \cite{35}:
\begin{equation}\label{eq28}
F=-\frac{\dot{M}}{4\pi\sqrt{-g}}\frac{\Omega_{,r}}{(E-\Omega L)^2}\int_{r_{in}}^r(E-\Omega L)L_{,r}dr,
\end{equation}
Where $\dot{M}$ denotes the mass accretion rate, $g$ represents the metric determinant, and $r_{\text{in}}$ signifies the radius of the inner edge of the accretion disk. $E$, $L$, and $\Omega$ respectively stand for the energy, angular momentum, and angular velocity of the particle in circular orbit.

\par
It's important to note that while the Novikov-Thorne model is originally formulated for Kerr BHs, several studies have adapted this model for spherically symmetric black holes \cite{36,37,38,39,40}. Only when $g_{t\varphi}=0$ for spherically symmetric black holes, modifications are required for the expressions of $E$, $L$, and $\Omega$. Under the metric of spherical symmetry $ds^2=g_{tt}dt^2+g_{rr}dr^2+g_{\theta \theta}d\theta^2+g_{\varphi \varphi}d\varphi^2$, the energy, angular momentum, and angular velocity of particles on the equatorial plane are respectively:
\begin{equation}\label{eq29}
E=-\frac{g_{tt}}{\sqrt{-g_{tt}-g_{\varphi \varphi}\Omega^2}},
\end{equation}
\begin{equation}\label{eq30}
L=\frac{g_{\varphi \varphi}\Omega}{\sqrt{-g_{tt}-g_{\varphi \varphi}\Omega^2}},
\end{equation}
\begin{equation}\label{eq31}
\Omega=\frac{d\varphi}{dt}=\sqrt{-\frac{g_{tt,r}}{g_{\varphi \varphi,r}}}.
\end{equation}
\begin{figure*}
\centering
\begin{minipage}{1\textwidth}
\centering
\includegraphics[width=1\textwidth]{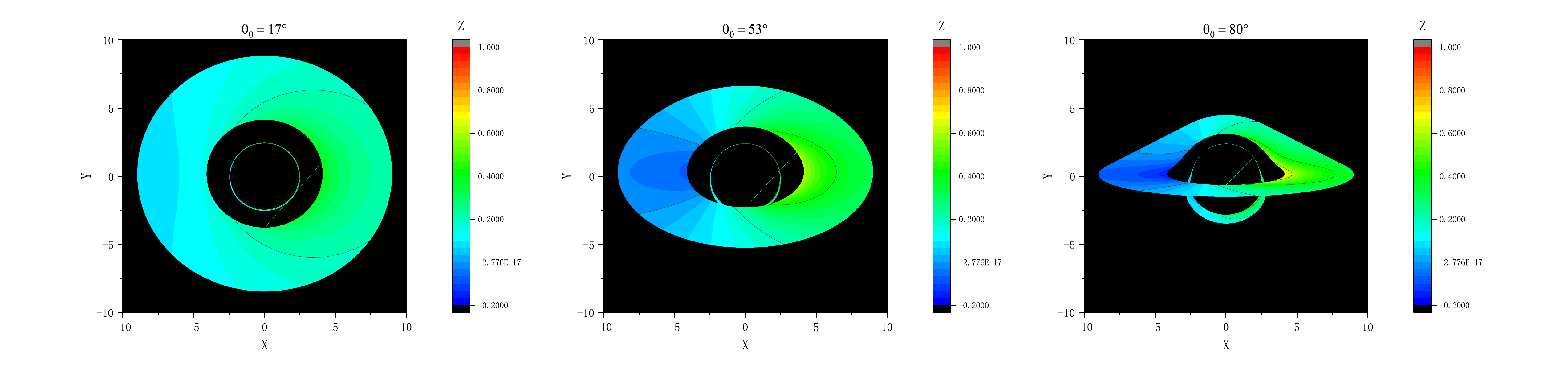}
\end{minipage}
\begin{minipage}{1\textwidth}
\centering
\includegraphics[width=1\textwidth]{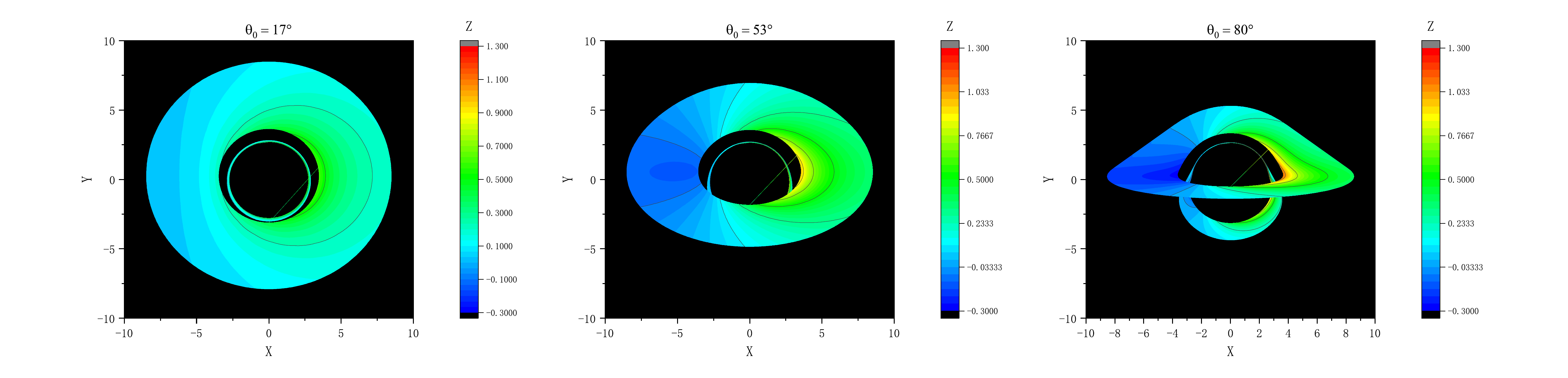}
\end{minipage}
 \caption{Redshift distribution of the accretion disk of two types of BHs with the viewing angles $\theta_0=17^\circ$, $53^\circ$ and $80^\circ$. Top panel: Non-Schwarzschild BH. Bottom panel: Schwarzschild BH. }
\label{t9}
\end{figure*}

\par
To simplify calculations, we utilize the maximum radiation emitted by the Schwarzschild BH as the unit of radiation flux. Fig. \ref{t8} illustrates the radiation flux distribution of two types of BHs at different viewing angles. It's observable that the radiation from the accretion disk approaches zero at the inner edge of the disk, and the radiation flux initially rises and then decreases with increasing radius. Although the radiation distribution patterns of the non-Schwarzschild BH align with those of the Schwarzschild BH, the corresponding radiation values are smaller for the non-Schwarzschild BH.
	
\par
For observers positioned at infinity, both gravitational redshift and Doppler shift effects influence the observed radiation flux. As per the findings presented in \cite{51}, the observed radiation flux corresponds to the ratio of the radiation flux to the fourth power of the redshift factor $(z+1)$.
\begin{equation}\label{eq32}
F_{obs}=\frac{F}{(z+1)^4}.
\end{equation}
	
\begin{figure*}[htbp]
\centering
\includegraphics[width=18.2cm,height=10cm]{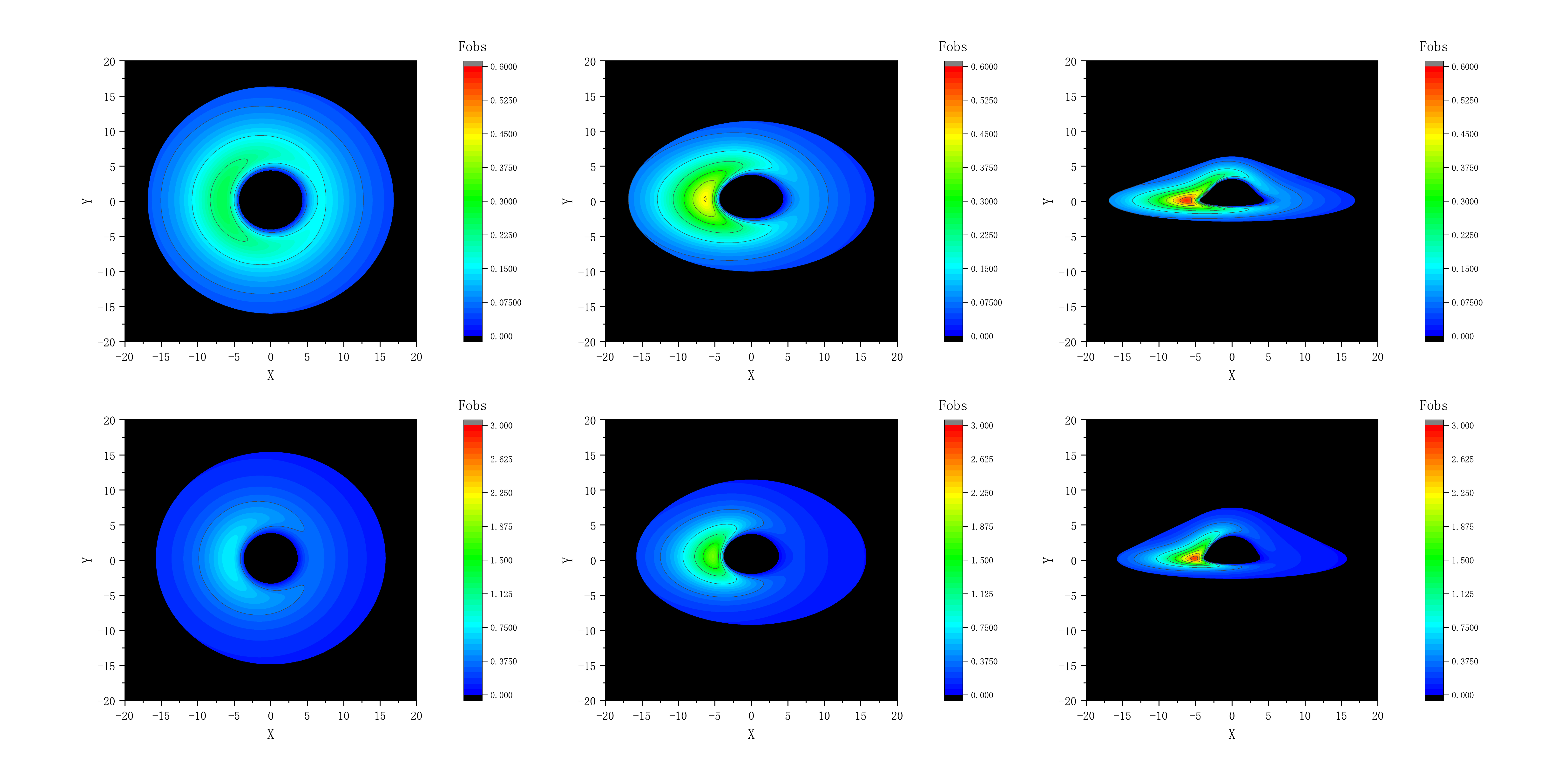}
\caption {The observed flux distribution of the dircet images of two types of BHs with the viewing angles $\theta_0=17^\circ$, $53^\circ$ and $80^\circ$. Top panel: Non-Schwarzschild BH. Bottom panel: Schwarzschild BH.}
\label{t10}
\end{figure*}
	
\par
The redshift factor is defined by the ratio of the energy $E_{\text{em}}$ at photon emission to the energy $E_{\text{obs}}$ observed at infinity. The energy at photon emission can be expressed in terms of the momentum of the photon $p$ and the velocity of the emitting particle $u$:
\begin{equation}\label{eq33}
E_{em}=p_tu^t+p_{\varphi}u^\varphi=p_tu^t(1+\Omega\frac{p_\varphi}{p_t}).
\end{equation}
	
\par
For a distant observer, the ratio $p_t/p_\varphi$ precisely represents the impact parameter of the photon at infinity with respect to the $z$ axis. By combining Eq. (\ref{eq29}) and Eq. (\ref{eq33}), one can derive the expression for the redshift factor:
\begin{equation}\label{eq34}
1+z=\frac{E_{em}}{E_{obs}}=\frac{1+b\Omega\sin{\alpha}\sin{\theta_0}}{\sqrt{-g_{tt}- g_{\phi \phi}\Omega^2}}.
\end{equation}
\begin{figure*}[htbp]
\centering
\includegraphics[width=18.2cm,height=10cm]{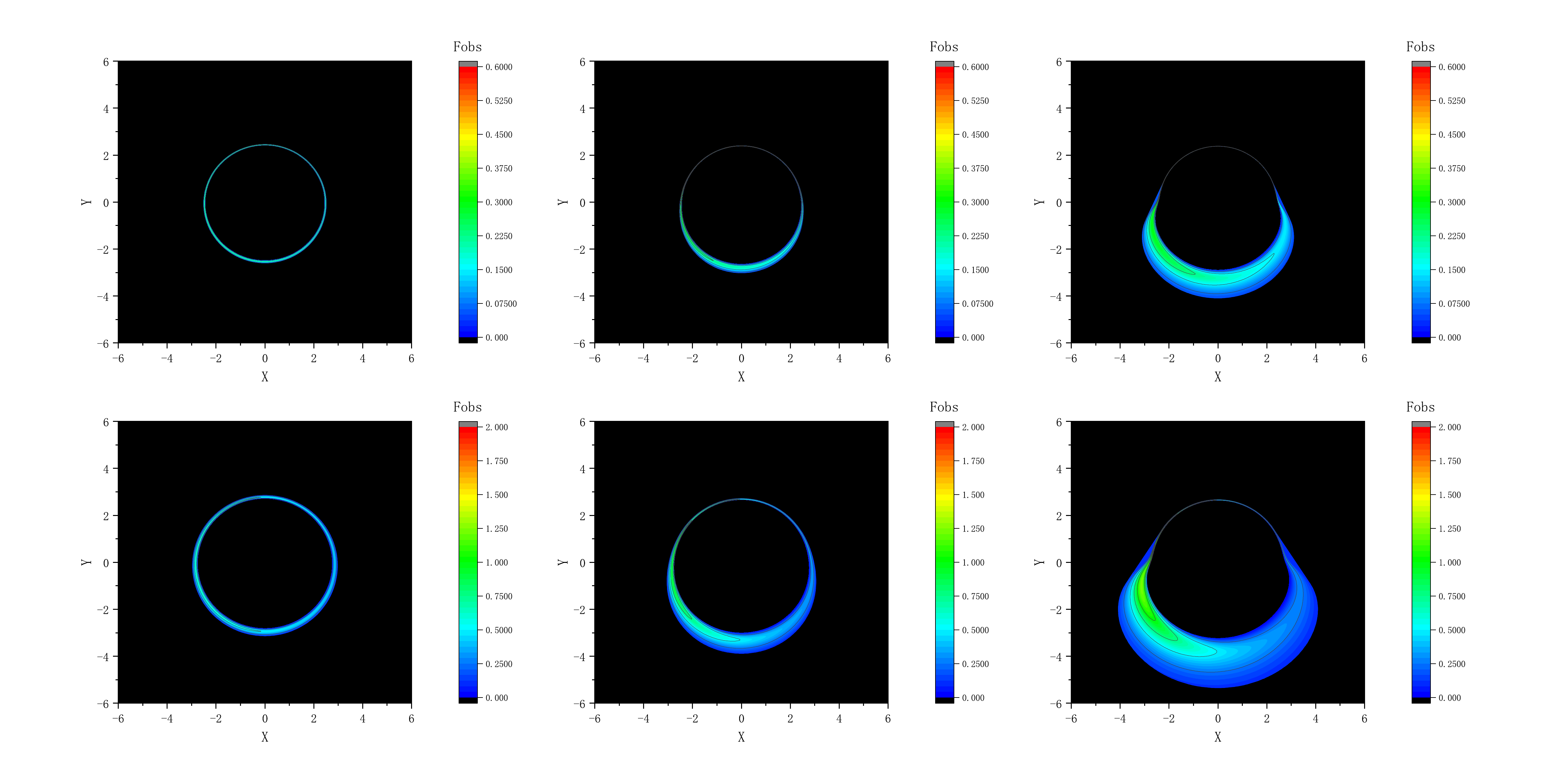}
\caption {The observed flux distribution of the secondary images of two types of BHs with the viewing angles $\theta_0=17^\circ$, $53^\circ$ and $80^\circ$. Top panel: Non-Schwarzschild BH. Bottom panel: Schwarzschild BH.}
\label{t11}
\end{figure*}
\begin{figure*}
\centering
\includegraphics[width=18.2cm,height=10cm]{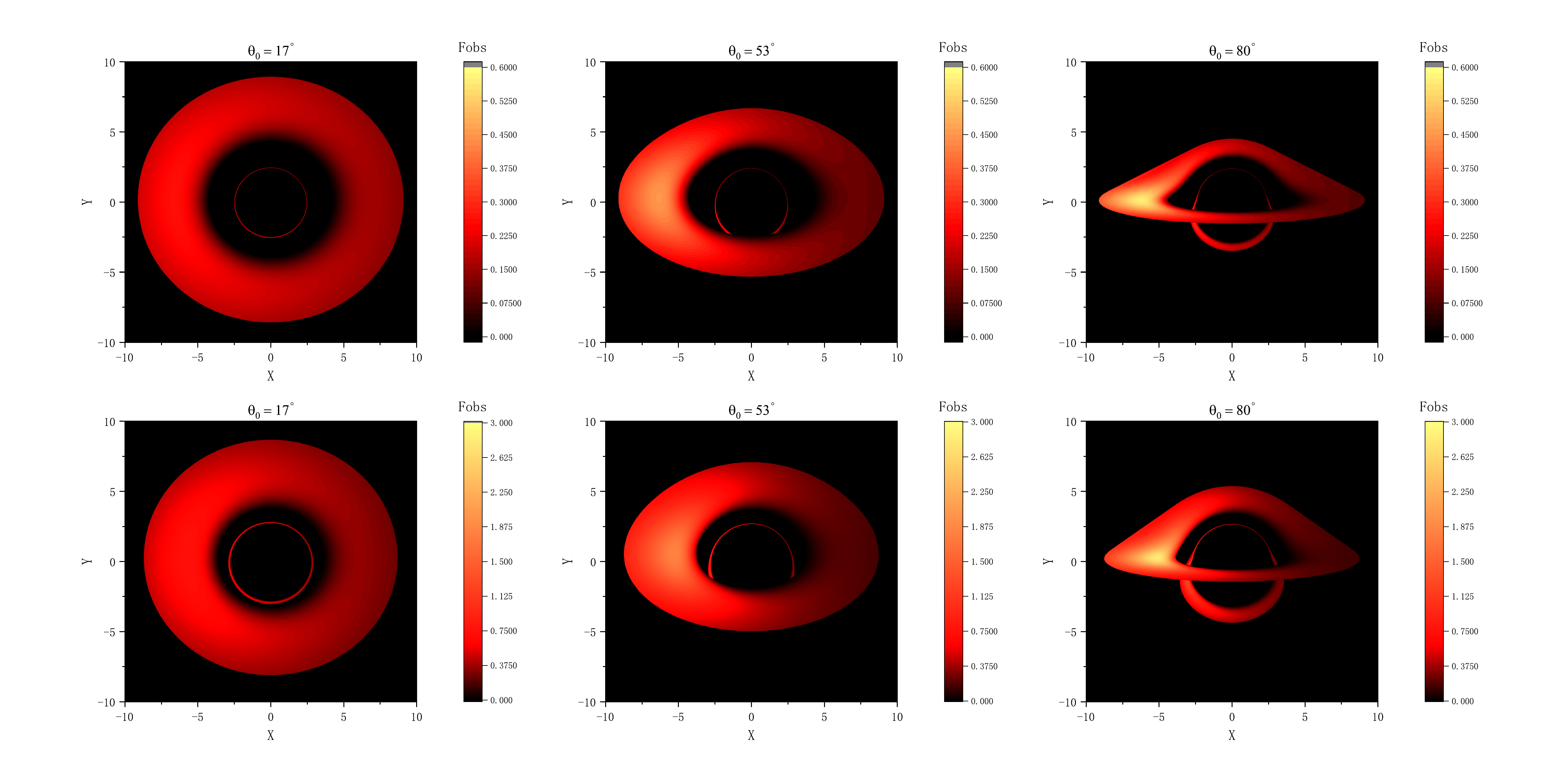}
\caption{Numerical simulation image of two types of BHs with the viewing angles $\theta_0=17^\circ$, $53^\circ$ and $80^\circ$. Top panel: Non-Schwarzschild BH. Bottom panel: Schwarzschild BH.}
\label{t12}
\end{figure*}

\par
Figure \ref{t9} displays the redshift distribution at various viewing angles. It's notable that, besides the symmetric gravitational redshift, the overall redshift distribution exhibits asymmetry due to the rotation of the accretion disk. On the side rotating towards the observer, the redshift is diminished, and may even shift towards blueshift, whereas on the other side, the redshift effect is amplified. Additionally, the viewing angle influences the redshift distribution, with a more pronounced redshift effect observed as the viewing angle increases. The redshift distribution patterns of the two types of black holes are fundamentally similar, although the effect is slightly weaker for non-Schwarzschild BHs.
	
\par
Considering both the radiation flux and redshift of the accretion disk, we can deduce the observed flux distribution of the accretion disk (refer to Fig. \ref{t10} and Fig. \ref{t11}). Subsequently, we simulate the images of the two types of BHs based on these observed flux distributions (Fig. \ref{t12}). It's evident that for both BHs with $r_0=1$, the images exhibit strong asymmetry, and the brightness distribution patterns are similar. However, notable differences arise: the non-Schwarzschild BHs appear weaker, and their secondary images are smaller compared to those of the Schwarzschild BHs. These distinctions could potentially serve as observable features for distinguishing between the two types of BHs.

\section{CONCLUSIONS}
\label{sec:5}
\par
In this analysis, we explore the optical characteristics of non-Schwarzschild BH surrounded by thin disks within higher derivative gravity, identifying observable features that distinguish them from Schwarzschild BHs. We replicate the investigations conducted in \cite{6}, obtaining solutions for non-Schwarzschild BHs with $r_0=1$, $r_0=2$, and $r_0=3$. Subsequent analysis of time-like geodesics reveals that the geodesic properties of non-Schwarzschild BHs with $r_0=1$ closely resemble those of Schwarzschild BHs, whereas non-Schwarzschild BHs with $r_0=2$ exhibit distinct characteristics. Notably, particles with any angular momentum, even zero angular momentum, must traverse a potential barrier to approach the BH from a distance, suggesting a effective repulsive force exerted by the BH in this scenario. To illustrate this property effectively, we present the unbound orbit diagram of particles with varying energies for the case $r_0=2$ (Fig. \ref{t3}), revealing that only particles with energies greater than the maximum potential energy $V_{\text{max}}$ can reach the BH. Furthermore, numerical results regarding the stability of circular orbits indicate that BHs with $r_0>r_c$ lack stable time-like circular orbits, precluding the formation of accretion disks around the BH. Consequently, observing such BHs may prove challenging.

\par
For the case of $r_0=1$, we conduct further analysis on null geodesics within the BH spacetime, calculating the photon sphere radius and the corresponding critical impact parameters of non-Schwarzschild BHs using the effective potential. Additionally, we employ numerical integration to determine the deflection angle of photons around Schwarzschild BHs, and plot the trajectories of photons around both Schwarzschild and non-Schwarzschild BHs (Fig. \ref{t4}). To visualize the BH images, we utilize the ray-tracing method proposed by Luminet \cite{36} to obtain the direct and secondary images of the black hole accretion disk (Fig. \ref{t7}). Our findings indicate that while there's no significant difference between the direct images of the two types of BHs, the degree of separation of secondary images for non-Schwarzschild BHs diminishes compared to Schwarzschild BHs as the radius increases. Furthermore, in line with previous research, the primary image of non-Schwarzschild BHs also exhibits a "hat" shape with increasing viewing angles, and the separation of secondary images becomes more apparent.

\par
Furthermore, we take into account the radiation emitted by the accretion disk and calculate the distribution of the radiation flux using the Novikov-Thorne accretion disk formula from \cite{35} (Fig. \ref{t8}). Our results indicate that while the distribution patterns of the radiation flux for both types of black holes are essentially the same, the radiation intensity of non-Schwarzschild BHs is comparatively weaker. Additionally, gravitational redshift and the Doppler effect influence the observed radiation flux. By combining the radiation flux distribution with the redshift distribution of the accretion disk, we calculate the observed radiation flux distribution (Fig. \ref{t10} and Fig. \ref{t11}). Finally, we conduct numerical simulations of the observed radiation flux to generate optical images of the two types of BHs (Fig. \ref{t12}). The results also reveal that the brightness distribution patterns of the two types of BHs are similar. However, due to the redshift effect, both types of images exhibit asymmetry, with the degree of asymmetry increasing with the viewing angle. Notably, we observe that non-Schwarzschild BHs appear dimmer, and the secondary images of non-Schwarzschild BHs are less separated compared to those of Schwarzschild BHs. These differences may serve as observable features to distinguish between the two types of BHs.

\begin{acknowledgments}
    This project was supported by National Natural Science Foundation of China (NNSFC) under contract No. 42230207 and the Fundamental Research Funds for the Central Universities, China University of Geosciences (Wuhan) with No. G1323523064.
\end{acknowledgments}

\end{document}